# The electronic disorder landscape of mixed halide perovskites

Yun Liu[1], Jean-Philippe Banon[2], Kyle Frohna[1], Yu-Hsien Chiang[1], Ganbaatar Tumen-Ulzii[3], Samuel D. Stranks[1,3], Marcel Filoche[2], Richard H. Friend[1*]

[1]Cavendish Laboratory, University of Cambridge, Cambridge, CB3 0HE, UK

[2]Laboratoire de Physique de la Matière Condensée, CNRS, Ecole Polytechnique, Institut Polytechnique de Paris, 91120 Palaiseau, France

[3]Department of Chemical Engineering & Biotechnology, University of Cambridge, Cambridge, CB3 0AS, UK

*Email: rhf10@cam.ac.uk

# Abstract

Bandgap tunability of lead mixed-halide perovskites makes them promising candidates for various applications in optoelectronics since they exhibit sharp optical absorption onsets despite the presence of disorder from halide alloying. Here we use localization landscape theory to reveal that the static disorder due to compositional alloying for iodide:bromide perovskite contributes at most 3 meV to the Urbach energy. Our modelling reveals that the reason for this small contribution is due to the small effective masses in perovskites, resulting in a natural length scale of around 20nm for the "effective confining potential" for electrons and holes, with short range potential fluctuations smoothed out. The increase in Urbach energy across the compositional range agrees well with our optical absorption measurements. We model systems of sizes up to 80 nm in three dimensions, allowing us to explore halide segregation, accurately reproducing the experimentally observed absorption spectra and demonstrating the scope of our method to model electronic structures on large length scales. Our results suggest that we should look beyond static contribution and focus on the dynamic temperature dependent contribution to the Urbach energy.

Mixed-halide perovskites (general formula ABX$_3$) have become the focus in the search for next generation solar cells and light-emitting diodes (LED)[1–3]. Their bandgap tunability across the visible and infrared spectrum coupled with a relatively clean bandgap, high-throughput synthesis conditions and high internal quantum yield make them attractive and promising for these applications[4,5]. For example, efficient infrared and blue LEDs have been based on mixed halide compositions MAPbI$_{3-x}$Cl$_x$ and Cs$_y$FA$_{1-y}$Pb(Br$_{1-x}$Cl$_x$)$_{3.5}$ [6,7] (FA, formamidinium; MA, methylammonium). The triple-cation lead halide perovskite (Cs$_{0.05}$FA$_{0.78}$MA$_{0.17}$)Pb(I$_{0.83}$Br$_{0.17}$)$_3$ is a highly reproducible and versatile system that shows good stability and photovoltaic efficiency[8]. For perovskite-silicon tandem cells, a Br fraction of up to 40% is used; the higher Br content being needed for the required bandgap[9]. Alloying in the halide sites in these high-performance devices are necessary to achieve the desired efficiency and structural stability[10,11].

The mixing of halides in perovskite inevitably introduces compositional disorder into the material, and the degree of disorder can be empirically characterized using the Urbach energy[12]. The Urbach energy can be measured from the absorption spectrum which shows a characteristic sharp band edge[13,14], below which there is a rapid fall off parametrized with the following exponential dependence

$$\alpha = \alpha_0 \exp\left(\frac{\hbar\omega}{E_U}\right) \quad (1)$$

where $\alpha$ is the absorption coefficient, $\alpha_0$ is a material constant, $\hbar\omega$ is the photon energy, and $E_U$ is the Urbach energy. The Urbach energy has been shown to change with halide composition in systems such as MAPb(Br$_{1-x}$I$_x$)$_3$[14]. This band tailing in the density of states and the absorption spectra reduces the radiative efficiency of solar cells below the Shockley-Queisser limit by deviating from the step-function absorption assumption[15–17].

Further reduction of radiative efficiency can also arise in mixed halide perovskite from photo-induced halide segregation that results in bandgap instability, reduced open circuit voltage and carrier mobility[18,19]. As first reported by Hoke *et al*[20], MAPb(I$_{1-x}$Br$_x$)$_3$ undergoes reversible halide phase segregation under illumination into I and Br rich domains. By comparing the peak of the red-shifted photoluminescence (PL) spectra and local lattice parameter from powder XRD measurements, the I-rich region has been indirectly inferred to have an approximate composition of $x = 0.2$, for any given starting halide composition with $x > 0.2$. The cause of the segregation is much debated with various microscopic models proposed to understand the phenomenon with the goal to minimize it[21–25]. The literature also disagrees on how much of the parent phase will turn I-rich under continuous illumination, with results varying from 1% to as much as 23%[20,26–28]. Therefore, we need a systematic and fundamental understanding on how halide segregation impacts the optical properties of mixed halide perovskites in order to understand and ultimately control this disorder.

Quantum mechanical modelling has played an important role in understanding the optical properties of mixed halide perovskites. Average properties such as lattice parameters, formation energies and bandgap can be predicted with sufficient accuracy based on density functional theory calculations using quasi-random models or large supercells[29–33]. On the other hand, there are only a few studies on absorption tails in semiconductors. The temperature dependent Urbach energy of amorphous silicon and silica glass were studied using *ab initio* molecular dynamics simulation[34,35]. A very dense **k**-point sampling of the Brillouin zone is used to compute the band tail states in many single phase semiconductors including MAPbI$_3$[36]. The high chemical complexity of perovskites, the large supercells needed to capture the effect of halide alloying and segregation, and the need to include quantum mechanical effects represents a huge challenge to understand the tail states arising from exponentially rare absorption events.

To overcome these difficulties, we utilize a recently developed theoretical framework called the "Localization Landscape (LL)[37]" to investigate the effects of disorder induced by halide composition and segregation. Briefly speaking, the right-hand side of the Schrödinger equation is replaced with a constant in the LL framework to arrive at the landscape equation

$$\widehat{H}u = \left(-\frac{\hbar^2}{2m^*}\nabla^2 + V\right)u = 1 \qquad (2)$$

where $\hbar$ is the reduced Planck constant, $m^*$ is the effective mass of electron or hole, $V(\mathbf{r})$ is the disordered potential. The localization landscape, $u(\mathbf{r})$, is a solution of the landscape equation with the appropriate boundary conditions and it contains a network of surfaces in 3D that generates an invisible partition of the system that identifies the subregions confining the quantum waves. The inverse of the landscape $W(\mathbf{r}) = 1/u(\mathbf{r})$ can be interpreted as a semi-classical effective confining potential determining the strength of the confinement as well as the long-range decay of the quantum states[38].

The LL theory is a very promising tool to understand quantum waves in disordered potentials, as it provides a speedup of two to three orders of magnitude compared to solving the Schrödinger equation self-consistently. The LL equation has been coupled with the classical drift-diffusion model to compute carrier transports in 2D alloyed nitride quantum wells, successfully reproducing the quantum tunnelling and quantum confinement effect[39,40].

Absorption models based on the LL equation have also accurately reproduced Urbach energy in disordered InGaN systems[41,42]. Analogous to the nitride system, the perovskites are direct bandgap semiconductors, and alloying at the halide sites offers stable phases across the full compositional range.

By modelling the system at a large length scale of 80 nm in three dimensions, we show that the maximum disorder arising from compositional alloying contributes 3 meV in the Urbach energy. This can be understood from the natural length of around 20 nm for the "effective confining potential" computed from the LL theory, with short range potential fluctuations in the alloys smoothed out. The 3 meV increase in Urbach energy in the studied compositional range is also in good agreement with our optical absorption measurements. The static disorder in both random alloy and halide segregated system is much smaller than the experimentally measured Urbach tail. Current indications are that this is dominated by thermal contributions[14,17,43].

**Random Alloy Modelling and Experiments**

We first focus on the widely employed triple cation perovskite $(Cs_{0.05}FA_{0.78}MA_{0.17})Pb(I_{0.83}Br_{0.17})_3$. The landscape equation uses an effective mass approximation with input parameters being the local electron and hole effective masses, and the conduction and valence band potentials. These values are computed from their pure phases data listed in Table S1[44–48]. Triple cation perovskite adopts a slightly tetragonal phase at ambient conditions and the electronic properties are largely inherited from the parent FA system[49]. As the tilting of the $PbI_6$ octahedral is small, we use the cubic phase in this study (Figure 1a). There are many reports on the lattice parameter and bandgap for cubic FA perovskites, and those values used in this work are a good representation of these published results. For effective masses and Kane's energy, we used the value for the Cs based perovskites, since the A site cations do not directly participate in the bonding and only influence the electronic structure indirectly by changing the lattice parameter and structural stability[50,51]. The A site cations are therefore also not included in the parametrization of the landscape model and only the halides are considered. We also ignore the effect of the anisotropy introduced by the FA cations as they produce a well-behaved effective electronic structure from their average positions[52].

The local potentials for hole and electron are determined from the bandgaps ($E_g$), and the energetic alignments of the valence band maximum (VBM) and the conduction band minimum (CBM) of the pure species, characterized by the parameter γ (details in Methods section). Due to the undetermined nature of the band alignments reported in the literature[46,53], we use 2 schemes to represent the two boundary cases (Figure 1b). In Scheme A with γ = 0.5, the band offset is equally distributed between the CBM and VBM. In Scheme B, the band offset is entirely attributed to the VBM with the CBM energy level being completely flat (γ = 0). Another important parameter to consider is the Gaussian smearing parameter $\sigma$ (Equation 3 in Methods section) which defines the maximum length scale over which the rapidly fluctuating distribution of atoms can be averaged to obtain a continuous potential while preserving the effects of disorder on the electronic properties. A minimum smearing parameter of $\sigma = \frac{a}{2}$ is necessary to capture the nearest-neighbor halide environment in a unit cell, which was used as a starting value.

We generate the random alloy by distributing halide atoms on the sites of the halide sublattice, assuming a spatial uniform probability equal to the mean halide composition. The simulation box is a cube of side length of 80 nm, containing approximately 2.05 million unit cells and 6.15 million halide atoms. We then computed the local potential (details in Methods section). A 2D cut of the 3D input valence band potential $E_v(\mathbf{r})$ using alignment scheme B is shown in Figure 1c, exhibiting random energetic fluctuations on the orders of ~500 meV. The valence and the conduction potentials, $E_v(\mathbf{r})$ and $E_c(\mathbf{r})$ using alignment scheme A exhibits the same randomness but with a smaller energetic fluctuation (Supplementary Figure S1).

For the standard triple cation composition with Br:I = 17:83, the resulting effective confining potential for holes $W_v(\mathbf{r})$ computed using the landscape equation is plotted in Figure 1d along the same 2D plane as $E_v(\mathbf{r})$. We note that the fluctuations in the potential have been significantly reduced to be around ~40 meV. The typical spatial regions of the valleys in $W_v(\mathbf{r})$ also increase to be ~20 nm, much larger than those of the input potential. As $W_v(\mathbf{r})$ acts as an effective confining potential seen by the hole eigenstates, short range disorder smaller than the length scale of 20 nm is smoothed out. We extended our analysis to higher Br concentrations at $x = 0.33$ and $x = 0.50$, corresponding to typical composition for perovskite-silicon tandem cell and maximum geometric disorder (Figure S2). Across this full compositional range, the spatial fluctuations in $W_v(\mathbf{r})$ that governs the nature of the hole states at the top of the valence band, are very much smaller than for the input potential $E_v(\mathbf{r})$.

To quantify the degree of disorder due to the effective confining potential, we computed absorption spectra (α) using the Wigner-Weyl approximation. This approach developed by Banon *et. al.*[42] consists of an exact reformulation in phase space of the absorption coefficient, given by Fermi's golden rule, *via* quasi-densities of states in phase space associated with the conduction and the valence band. The quasi-densities of states in phase space are approximated using the landscape-based modified Weyl law[38] leading to the closed form approximation of the absorption coefficient given in Equation 9 of the Methods. The final expression of the absorption coefficient takes the form of a spatial average of a local absorption coefficient which one would obtain for a homogeneous direct band gap semiconductor but with a fluctuating effective band gap. Figure 2a displays computations of the absorption coefficient spectra for different Br concentrations. Above the average $E_g$, the absorption coefficient follows a general square root behaviour as expected from direct bandgap 3D semiconductors. Below $E_g$, there is a distinct tail from which we can fit exponential curves to extract the Urbach energies. For our choice of band alignment $\gamma = 0$ and Gaussian smearing parameter $\sigma = 0.5a$, $E_U$ increases from approximately 1.2 meV to 3.0 meV as the Br concentration is increased from 0.17 to 0.50. Setting $\gamma = 0.5$ lowers the input potential barrier for the quantum states that results in more smeared out effective potentials, resulting in smaller $E_U$ (Figure 2b). Increasing the Gaussian smearing parameter also reduces $E_U$, which is expected since increasing $\sigma$ means decreasing amplitude of the conduction and valence potential (Figure S3).

The choice of $\gamma = 0$ and $\sigma = 0.5a$ sets the maximum amount of static disorder in the system within our model, and the corresponding maximum value for the Urbach energy in mixed bromide-iodide perovskites is 3 meV. We attribute the small compositional disorder to the small hole effective masses in perovskites (0.095 $m_0$ and 0.128 $m_0$ for I and Br species respectively). We note that this is considerably smaller that for III-V nitrides such as InGaN at around 20 meV, where effective masses of the heavy holes are considerably larger ($m_h$ = 1.87 $m_0$ and 1.61 $m_0$ for GaN and InN respectively).

To further understand the nature of the composition dependent Urbach energies, we synthesized perovskite films of compositions $(Cs_{0.05}FA_{0.78}MA_{0.17})Pb(I_{1-x}Br_x)_3$ and performed photothermal deflection spectroscopy (PDS) measurements. PDS is an ultrasensitive absorption measurement technique, which allows detection and quantification of sub-bandgap features for thin films (details in Methods). Figure 2c shows the PDS spectra for samples at the same compositions of $x = 0, 0.17, 0.33, 0.50$. The PDS spectra exhibit a sharp sub-bandgap decay indicating a clean bandgap that is characteristic of perovskites. The film processing of the pure iodide composition with triple cation A site, i.e. $(Cs_{0.05}FA_{0.78}MA_{0.17})PbI_3$, is not optimized so we used the pure FA counterpart $FAPbI_3$ instead.

Figure 2d shows the extracted Urbach energies, and they are in broad agreement with previously reported values in the literature[13,14,43,54,55]. The values of the measured Urbach energies are all above 10 meV, much larger than our modelled values. This discrepancy has been attributed to the strong temperature dependent contribution to the Urbach energy that has been previously noted and assigned to dynamic disorder due to thermal occupancy of vibrational modes[43,55]. Some studies using temperature dependent photoluminescence measurements have indicated that extrapolation to low temperatures takes the Urbach energies to a few meV[56]. We note that our direct measurements of optical absorption are performed at room temperature, and we emphasise that our computational model introduces disorder only through the compositional alloying of the halides.

The pure iodide composition ($x = 0$) has an Urbach energy of around 15.5 meV as shown in the Figure 2d. Adding compositional disorder does not change the Urbach energy much. We note that this rises by a few meV for the $x = 0.33$ and 0.5 compositions and is, surprisingly, lower at $x = 0.17$, down to 13 meV. This is probably due to the much-improved film homogeneity and stability at $x = 0.17$, as recipes are optimised for this specific composition[8]. Our results illustrate that compositional alloying contributes only a small amount of disorder, in excellent agreement with our modelling. It appears that the largest source of disorder in perovskites comes from the dynamic temperature dependent component.

**Halide Segregation**

Under illumination, halide segregation typically occurs at $x > 0.2$, and a I-rich region emerges that dominates the PL spectra. While highly emissive[57,58], these regions will result in lower carrier mobilities and reduction in open circuit voltages due to the lower effective bandgap[54]. While several empirical approaches such as anion treatment[59], cation engineering[60] and compressive stress[61] have been proposed to limit the degree of segregation, there still lacks an atomistic understanding on the I-rich region and its impact on the electronic structure of the entire perovskite film. This is because typical experimental techniques such as PL are insensitive to the size and composition of the I-rich region, and X-ray diffraction is not quantitatively reliable due to the sensitivity to the sample texture[62]. Typical electronic structure methods also cannot handle the length scale at which halide segregation occurs.

Here we applied the same computational framework to a "sphere-in-a-box" model to investigate the impact of halide segregation on the optoelectronic properties of perovskites. Figure 4a shows the 2D cut of the 3D $E_v(\mathbf{r})$ of a single spherical I-rich region of diameter $d = 20$ nm embedded into a Br-rich majority phase of side length of 80 nm. The overall composition of the box is fixed at $x = 0.33$, and the I-rich region has $x = 0.2$ in line with the experimentally observed composition. Within each region, a random distribution of Br and I atoms is assumed. Figure 4b shows the computed $W_v(\mathbf{r})$ with the minority I-rich phase clearly providing a localized high energy region that are attractive for the hole states. The size of this region is approximately the same as the input I-rich region, with energetic fluctuations consistent with their random alloy counterpart.

We then generated I-rich regions of various sizes and compute their absorption spectra which are shown in Figure 4c. We first note that the effect of quantum confinement is observed for regions with $d \leq 20$ nm, whereby the absorption edges are blue-shifted. When we varied either the composition of the I-rich region (Figure S4a), or the overall halide composition (Figure S4b), quantum confinement seems to always occur at this length scale (Figure S3). In fact, we have observed that the hole effective potential of the random alloy fluctuates on this length scale in Figure 1d. As our model only takes into consideration the electron and hole effective masses and band edge energy offset, this universal length scale is therefore a fundamental material property of perovskites. Coincidentally, this length scale is approximately twice the exciton Bohr radius obtained for the Br and I perovskite systems at around 7.5-12.5 nm[47,63], even though the Coulomb interactions between electrons and holes are not included our model. In addition, we do note that discrete quantum dot transitions are not captured by our model, which only gives an effective overall trend that depends on the volume ratio of the majority and minority phases.

The overall shapes of the absorption spectra in Figure 4c also closely resemble the experimentally measured spectra for segregated perovskites using PDS or photocurrent spectroscopy[54,57], with a bump in the tail state due to the absorption of the minority I-rich phase. The $E_U$ fitted to the minority tail state is the same as that of the random alloy of the same composition (i.e. $x = 0.2$). This is because tail states are dominated by the fundamental states of the system, which are likely to be localized within the large attractive potential well within the I-rich region. We found that having additional I-rich spheres in the simulation box has negligible impact on the overall absorption spectra. (Figure S5).

Next, we focus on the "tail" of the majority region that arises due to the bump in the absorption spectra (Figure 5a). This tail has been attributed to an exponential distribution of bandgaps in the majority region, with the extracted $E_U$ significantly larger than that of the random alloy or the minority I-rich region[54]. To understand its origins, we plot the absorption power density of the segregated halide system, which gives the spatially resolved absorption at any given photon energy (Equation 10). At $\hbar\omega = 1.66$ eV, below the gap of the minority I-rich region, the random fluctuation of the iodide-rich alloy is contributing to the tail states, with no contributions from the majority phase (Figure 5b). When the photon energy increases from 1.68 eV to 1.72 eV, more states from within the I-rich region are contributing to the absorption (Figure 5c and 5d). As we approach the "tail" state of the majority region at $\hbar\omega = 1.74$ eV, we observe pockets of spatially separated regions in the majority region emerging, due to compositional disorder in the alloy (Figure 5e). There is also significant absorption coming from the I-rich region, since the photon energy is larger than its local bandgap, and its density of states is modelled here as the square root of the energy difference between the effective gap and the photon energy. The resultant "tail" of the majority region, therefore, is the sum of the optical contribution from both the minority and majority region. Our results show that we need to be careful when comparing the disorder of the I-rich minority region and majority region, as the higher value of the exponential fit does not necessarily represent higher degree of disorder. Finally, at $\hbar\omega = 1.76$ eV, both regions are making significant contribution to the absorption spectra as expected (Figure 5f).

**Conclusion**

In conclusion, we have applied the localization landscape theory together with the Wigner-Weyl approximation to absorption, to investigate the electronic and optical properties of mixed halide perovskites with significant computational speedup. We found that the Urbach energy due to compositional alloying in perovskites is around 3 meV, about an order of magnitude smaller than those in the InGaN systems. We show that this small disorder is due to the small electron and hole effective masses in perovskites, so that the electron and hole states "see" an effective confining potential with a natural length scale of around 20 nm with short range potential fluctuations in the alloys smeared out. The increase in Urbach energy across the compositional range agrees well with our optical absorption measurements. We have also reproduced the absorption spectrum of perovskites with halide segregation and observed quantum confinement effects for I-rich minority region smaller than the natural length scale of 20 nm. Our results show that static compositional disorder contributes a small amount of total disorder in perovskites, and we should focus on understanding the dynamic temperature dependent contribution to the Urbach energy arising from the electron-phonon interactions.

## Methods

**Local materials properties**

The lattice parameter *a* of the alloy is computed as the weighted arithmetic mean of the lattice parameter of the constituent phases[64]. The local averaged alloy composition $X(\mathbf{r})$ at any position **r** within the lattice is determined from the Gaussian averaging method with periodic boundary condition

$$X(\mathbf{r}) = \frac{\sum_i X_i e^{-\frac{(\mathbf{r}_i - \mathbf{r})^2}{2\sigma^2}}}{\sum_i e^{-\frac{(\mathbf{r}_i - \mathbf{r})^2}{2\sigma^2}}} \quad (3)$$

where the sum goes over all halide sites *i* of the supercell, $X_i = 1$ if the site is occupied by Br atom and $X_i = 0$ for iodide, and σ is the Gaussian smearing parameter. We use $\sigma = \frac{a}{2}, a$ and $2a$ in the current study.

After obtaining the local halide composition $X(\mathbf{r})$, $E_g(\mathbf{r})$, $E_c(\mathbf{r})$ and $E_v(\mathbf{r})$ are computed according to the following equations, as the band bowing parameter is negligible in the FA perovskites[64].

$$E_g^{\text{FAPb}(I_{1-x}Br_x)_3}(\mathbf{r}) = (1 - X(\mathbf{r}))E_g^{\text{FAPbI}_3} + X(\mathbf{r})E_g^{\text{FAPbBr}_3} \quad (4)$$

$$E_c^{\text{FAPb}(I_{1-x}Br_x)_3}(\mathbf{r}) = E_g^{\text{FAPbBr}_3} - \gamma\left(E_g^{\text{FAPbBr}_3} - E_g^{\text{FAPb}(I_{1-x}Br_x)_3}(\mathbf{r})\right) \quad (5)$$

$$E_v^{\text{FAPb}(I_{1-x}\text{Br}_x)_3}(\mathbf{r}) = (1-\gamma)\left(E_g^{\text{FAPbBr}_3} - E_g^{\text{FAPb}(I_{1-x}\text{Br}_x)_3}(\mathbf{r})\right) \quad (6)$$

where $\gamma$ is the band alignment parameter with $\gamma = 0.5$ and $\gamma = 0$ corresponding the Scheme A and B, respectively.

The local electron and hole effective masses are computed as follows

$$m_e^{\text{FAPb}(I_{1-x}\text{Br}_x)_3}(\mathbf{r}) = [(1-X(\mathbf{r}))/m_e^{\text{FAPbI}_3} + X(\mathbf{r})/m_e^{\text{FAPbBr}_3}]^{-1} \quad (7)$$

$$m_h^{\text{FAPb}(I_{1-x}\text{Br}_x)_3}(\mathbf{r}) = [(1-X(\mathbf{r}))/m_h^{\text{FAPbI}_3} + X(\mathbf{r})/m_h^{\text{FAPbBr}_3}]^{-1} \quad (8)$$

**Finite element computation**

The landscapes for the hole ($u_h(\mathbf{r})$) and electron ($u_e(\mathbf{r})$) are computed independently from Equation 2 using the finite element method with periodic boundary conditions. Meshes are generated using Gmsh[65] and we have used the finite element solver GetDP[66]. The size of the 3-dimensional domain studied has length $L = 80$nm, and the typical finite element step size is $\Delta x = 0.3$nm. The band edge data are interpolated on the nodal points. The discretized linear system is solved by using the iterative method of generalized minimal residual.

**Wigner-Weyl law for absorption**

The absorption spectra are computed using the Wigner-Weyl description of light absorption together with the LL theory, the validity and accuracy which has been systematically discussed in [42]:

$$\alpha(\omega) = \frac{e^2 E_p}{2\pi m_0 \varepsilon_0 c_0 \omega n(\omega)|\Omega|} \int_\Omega \left[\frac{2m_r(\mathbf{r})}{\hbar^2}\right]^{3/2} \left(\hbar\omega - W_g(\mathbf{r})\right)_+^{1/2} dr^3 \qquad (9)$$

where $e$ is the elementary charge, $E_p$ is the Kane's energy, $m_0$ is the rest mass of the electron, $\varepsilon_0$ is the vacuum permittivity, $c_0$ is the speed of light in vacuum, $n(\omega)$ is the real part of the refractive index, $\Omega$ is the volume of the system, $m_r$ is the spatially dependent reduced mass given by $\frac{1}{m_r(\mathbf{r})} = \frac{1}{m_e(\mathbf{r})} + \frac{1}{m_h(\mathbf{r})}$ and $W_g(\mathbf{r}) = W_c(\mathbf{r}) - W_v(\mathbf{r})$ is the effective bandgap, the + subscript denotes the positive part of the function. The effective potential is defined as the reciprocity of the respective landscape, $W_c(\mathbf{r}) = \frac{1}{u_c(\mathbf{r})}$ and $W_v(\mathbf{r}) = \frac{1}{u_v(\mathbf{r})}$.

The spatially dependent power absorption density, $\wp$ is given by the integrand within Equation 9 scaled by the length of the system $L$.

$$\wp(\mathbf{r}, \omega) = \frac{e^2 E_p}{2\pi m_0 \varepsilon_0 c_0 \omega n(\omega) S} \left[\frac{2m_r(\mathbf{r})}{\hbar^2}\right]^{3/2} \left(\hbar\omega - W_g(\mathbf{r})\right)_+^{1/2} \qquad (10)$$

where $S$ is the surface area of the system. For each absorption spectra, 50 independent realizations of the disordered alloy were run to obtain the average and the standard deviation.

**Perovskite synthesis**

Spectrosil quartz substrates were cleaned in a sonication bath for 15 minutes each in the following solvents: 1.5% v/v solution of Decon 90 detergent in deionised water, acetone the isopropanol. Then the substrates were surface treated with UV-ozone for 15 minutes before being transferred to a nitrogen filled glovebox with $O_2$ and $H_2O$ <1 ppm. Perovskite solutions were prepared with FAI (Greatcell Solar), methylammonium bromide (Greatcell Solar), $PbI_2$ (TCI) and $PbBr_2$ (TCI) in the relevant ratios were dissolved in anhydrous dimethylformamide/dimethyl sulfoxide (4:1 v/v, Sigma). 5% CsI (Sigma) was dissolved in dimethyl sulfoxide (1.5M) and then added to the precursor solution to finally prepare the perovskite triple cation solution with the relevant halide composition. 50 µl of perovskite solution was spread onto the quartz substrate and spin-coated in a two-step process, 1000 rpm for 10s then 6000 rpm for 20 s. 100 µl of chlorobenzene antisolvent was dropped onto the middle of the film 5s before the end of the second step. After the spin coating finished, the films were transferred to a hotplate at 100 °C and annealed for 1 hour.

Pure $FAPbI_3$ precursor solution was prepared by dissolving FAI (0.172 g, 1.0 mmol) and PbI2 (0.461 g, 1.0 mmol) in 1 ml DMF:DMSO (1:1 in volume). Additionally, 10 µl of ethylenediaminetetraacetic (EDTA, anhydrous, ≥ 99%) solution was added to the precursor solution as a stabiliser from the ETDA stock solution (0.01538 g/ml in DMSO). The concentration of EDTA is approximately 0.05 mol %. The precursor solution was stirred at 70 °C for 1 h and then filtered with a 0.2 µm polytetrafluoroethylene (PTFE) filter prior to use. The precursor solution was spin-coated on a pre-cleaned and UV-ozone treated quartz substrate at 1,000 rpm for 10 sec and then at 6,000 rpm for 30 sec. Five seconds before the end of the spinning program, 100 µl of antisolvent of chlorobenzene was dropped onto the spinning substrate. Subsequently, the perovskite layer was annealed at 150 °C for 30 min to form $FAPbI_3$[49].

**Photothermal Deflection Spectroscopy**

Perovskite films deposited on quartz substrates were submerged in Fluorinert FC-72, a chemically inert, optically transparent liquid medium. A xenon lamp was monochromated by a CVI DK240 monochromator before being modulated with a mechanical chopper at 13 Hz. The modulated light source was coupled to the film at normal incidence. A Qioptiq continuous wave, 670 nm laser was used to generate the probe beam is fibre coupled parallel to the front surface of the film. When light was absorbed by the film and generates heat, a refractive index gradient was generated in the liquid which deflected the probe beam. This deviation was modulated by the mechanical chopper and is detected by a photodiode then amplified using a Stanford Research Systems SR830 lock-in amplifier. The signal was then compared to a reference photodiode to normalise for spectral variations in the intensity of the light-source to extract the absorption spectra.

**Acknowledgements**

Y.L. and R.H.F acknowledges the funding from Simons Foundation (Grant 601946). J.-P. B. and M. F. acknowledges the funding from Simons Foundation (Grant 601944). K.F. acknowledges a George and Lilian Schiff Studentship, Winton Studentship, the Engineering and Physical Sciences Research Council (EPSRC) studentship, Cambridge Trust Scholarship, and Robert Gardiner Scholarship. Y.H.C. acknowledges funding from the Cambridge Trust. The work has received funding from the European Research Council under the European Union's Horizon 2020 research and innovation programme (HYPERION - grant agreement number 756962). S.D.S. acknowledges funding from the Royal Society and Tata Group (UF150033). The authors acknowledge funding from the EPSRC (EP/W004445/1). The calculations were performed using resources provided by the Cambridge Service for Data Driven Discovery (CSD3) operated by the University of Cambridge Research Computing Service (www.csd3.cam.ac.uk), provided by Dell EMC and Intel using Tier-2 funding from the EPSRC (capital grant EP/T022159/1), and DiRAC funding from the Science and Technology Facilities Council (www.dirac.ac.uk).

**Author Contributions**

R.H.F conceived the project. Y.L performed all the modelling and analysis using the code developed by J.-P. B., under the supervision of R.H.F and M. F.. K. F performed the PDS measurements and Y. H. C and G. T.-U. prepared the perovskites samples with the supervision of S. D. S.. Y. L. wrote the first draft of the manuscript. All authors participated in the process of revising the manuscript.

**Competing interests**

The authors declare no competing financial interest.

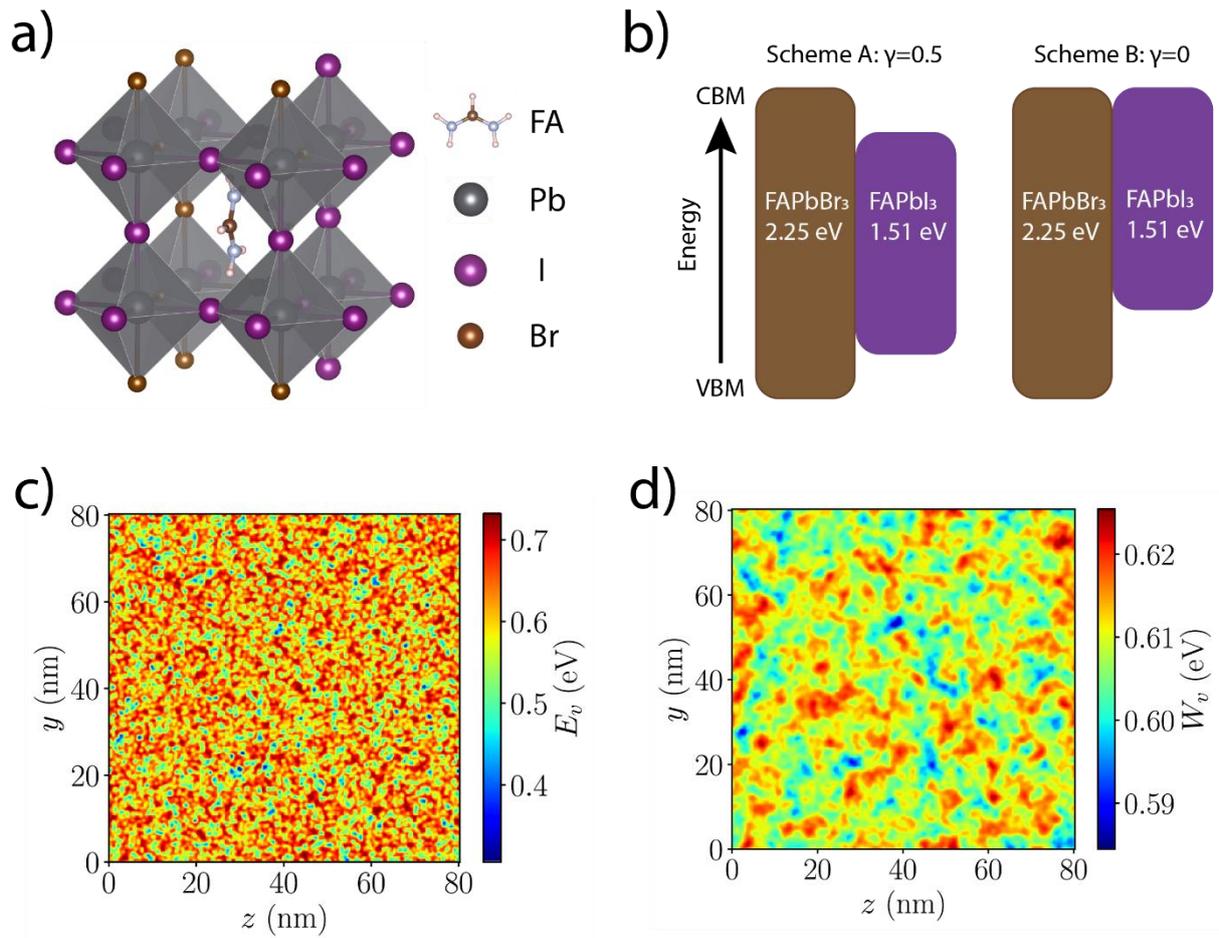

**Figure 1. Electronic properties of mixed halide perovskites** (a) A schematic visualization of the cubic phase of FA lead perovskite with random mix of halide atoms with the octahedral environment surrounding Pb shown. (b) The two different band alignment schemes used in this study. In Scheme A, the bandgap differences between halide species are evenly distributed between the CBM and VBM ($\gamma = 0.5$). In Scheme B, the offset is entirely attributed to the valence band, with the CBM levels perfectly aligned ($\gamma = 0$). (c) 2D cut of the local VBM energy $E_v(\mathbf{r})$ in the $yz$ plane for $FAPb(I_{0.83}Br_{0.17})_3$ whereby the I and Br atoms are randomly distributed. Band alignment scheme B is used with the energies referenced to the VBM of the pure Br system. (d) 2D cut of the computed effective confining potential $W_v(\mathbf{r})$ from the LL equation along the same $yz$ plane.

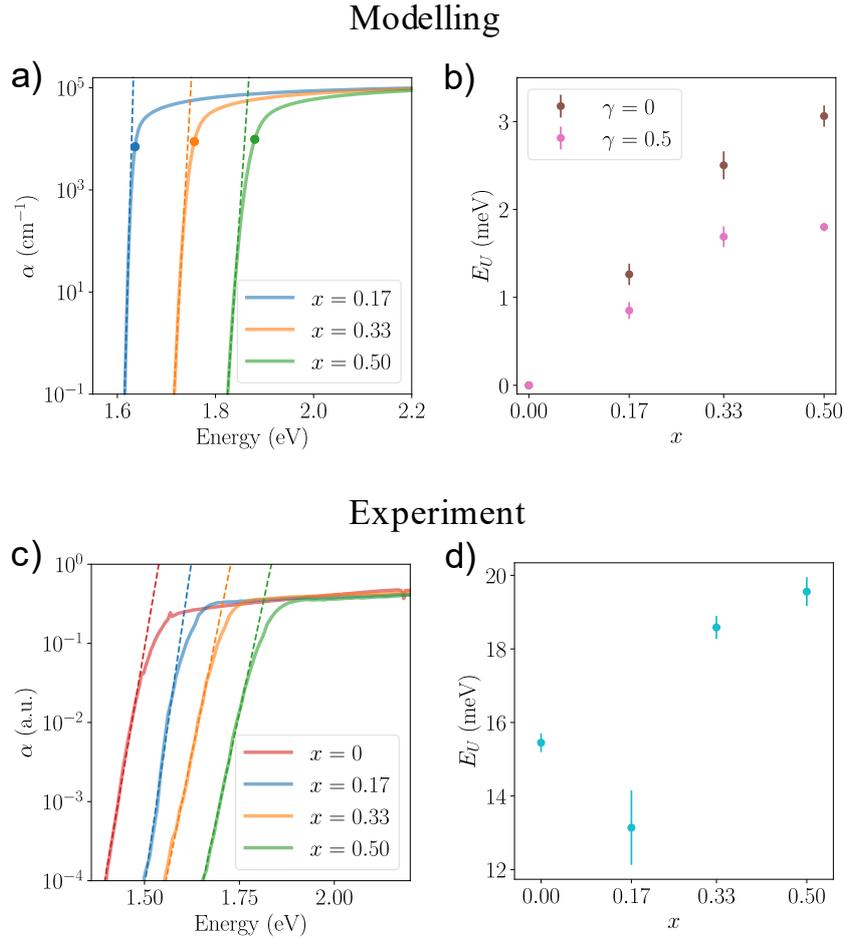

**Figure 2. Urbach tail of mixed halide perovskites from modelling and experiments.** (a) The computed absorption coefficient ($\alpha$) of FAPb(Br$_x$I$_{1-x}$)$_3$ in logarithmic scale as a function of the photon energy, averaged over 50 independent realizations. Solid circles indicate the average $E_g$ for each composition, and the dashed lines show the fitted exponentials below the absorption edges. Band alignment scheme B ($\gamma = 0$) and smearing parameter $\sigma = 0.5a$ are used. (b) The extracted Urbach energy as a function of the Br concentrations for both band alignment schemes using $\sigma = 0.5a$. (c) The measured absorption coefficient ($\alpha$) in logarithmic scale of FAPbI$_3$ and (Cs$_{0.05}$FA$_{0.78}$MA$_{0.17}$)Pb(I$_{1-x}$Br$_x$)$_3$ from photothermal deflection spectroscopy at room temperature. The dashed lines show the exponential fits below the absorption edges. (d) The extracted PDS Urbach energy as a function of Br compositions.

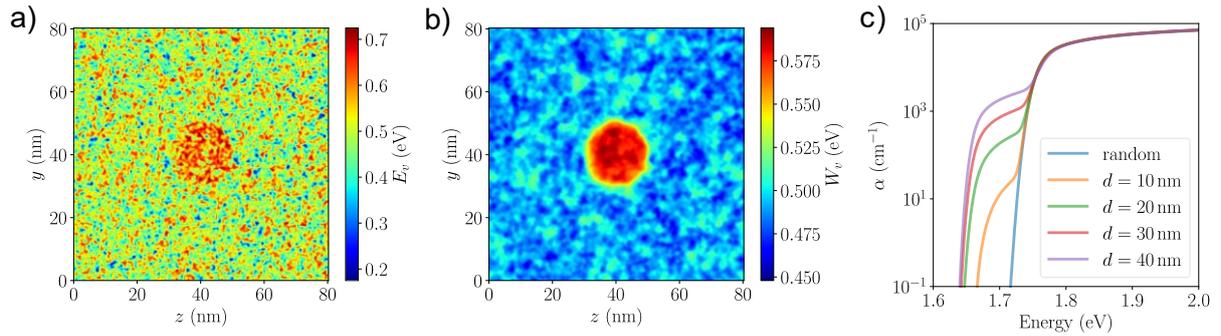

**Figure 3. Potentials and absorption spectra of halide segregated perovskites.** (a) 2D cut of the local $E_v(\mathbf{r})$ with a I-rich minority region ($x$=0.2) embedded into a majority phase with higher Br content, with the overall composition of the simulation box at $x$=0.33. The I-rich region has a diameter $d$=20 nm. (b) 2D cut of the effective potential $W_v(\mathbf{r})$ along the same plane. (c) The absorption spectra of the halide segregated perovskites shown in logarithmic scale, with I-rich regions of different sizes, averaged over 50 realizations. The overall composition of the simulation box is $x$=0.33 with I-rich regions having $x$=0.2. $d$ = 10 nm, 20 nm, 30 nm, 40 nm correspond to approximately 0.1%, 0.8%, 2.8% and 6.5% of the total volume of the perovskites turning I-rich. The absorption of the random alloy of FAPb(I$_{0.67}$Br$_{0.33}$)$_3$ is shown for comparison.

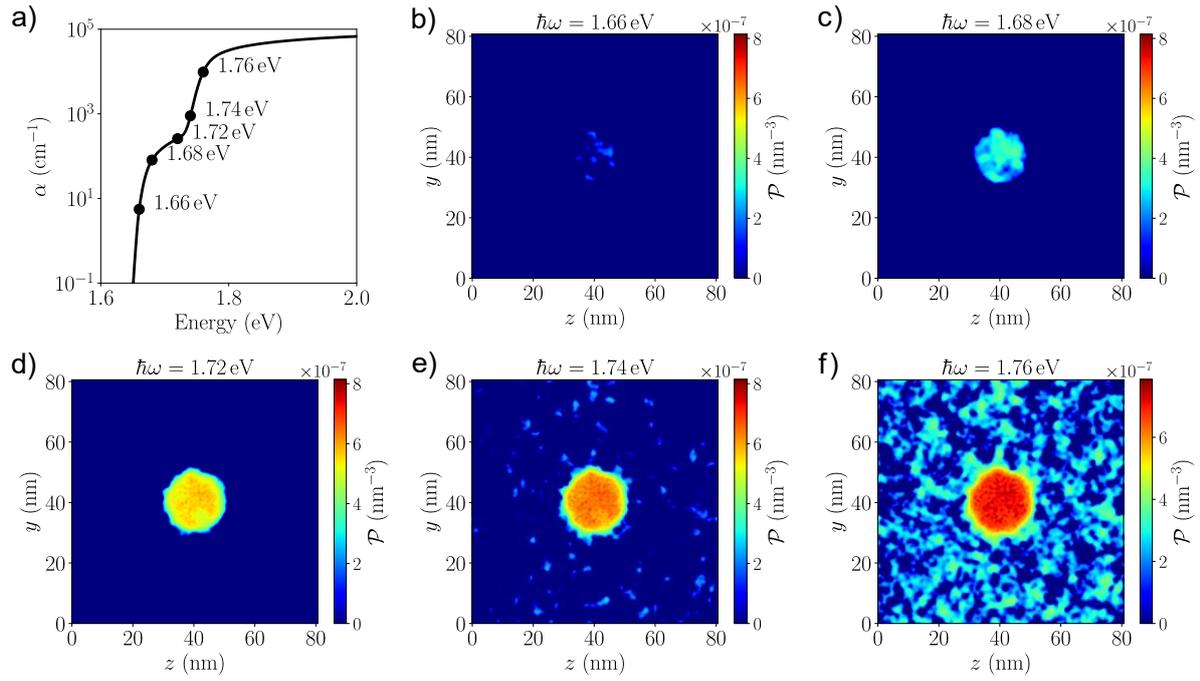

**Figure 4. Absorption power density of halide segregated perovskites.** (a) The absorption spectra of one realization of the halide segregated system, with the minority I-rich region having composition of $x = 0.20$, and the majority phase having composition of $x = 0.33$. (b)-(f) Absorption power density at increasing photon energies, going from 1.66 eV to 1.76 eV as labeled in (a).